\documentclass[journal=jacsat,manuscript=article]{achemso}
\setkeys{acs}{keywords = true}
\usepackage{chemformula} 
\usepackage[T1]{fontenc} 
\usepackage[english]{babel}
\usepackage[utf8x]{inputenc}
\usepackage{amsmath}
\usepackage{graphicx}
\usepackage[colorinlistoftodos]{todonotes}
\usepackage{gensymb}
\usepackage{textcomp}
\setkeys{acs}{articletitle = true}
\usepackage{natbib}
\usepackage{multirow}
\usepackage[flushleft]{threeparttable}
\usepackage{booktabs}
\usepackage{lipsum}
\usepackage[justification=centering]{caption}
\usepackage{setspace}
\SectionNumbersOn

\author{Yanhui Zhang}
\altaffiliation{School of Physics and CRANN, Trinity College, Dublin 2, Dublin, Ireland}
\email{yzhang1@tcd.ie}
\author{Stefano Sanvito}
\altaffiliation{School of Physics and CRANN, Trinity College, Dublin 2, Dublin, Ireland}
\email{sanvitos@tcd.ie}

\title[]{Interface engineering of graphene nanosheet reinforced ZrB$_2$ composites by tuning surface contacts}


\begin{document}

\begin{tocentry}
\includegraphics{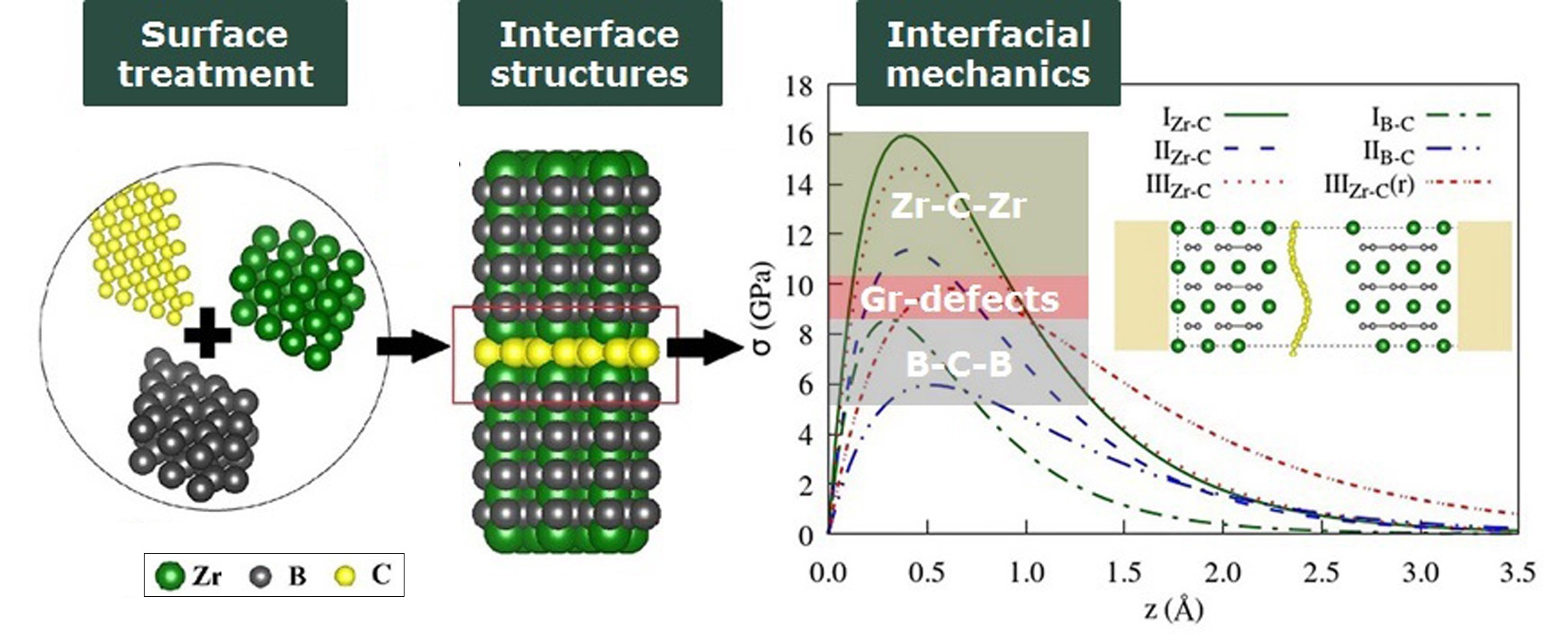}


\end{tocentry}

\begin{abstract}
The mechanical properties of heterophase interfaces are critically important for the behaviour of graphene-reinforced composites. 
In this work, the structure, adhesion, cleavage and sliding of heterophase interfaces, formed between a ZrB$_2$ matrix and graphene 
nanosheets, are systematically investigated by density functional theory, and compared to available experimental data. We 
demonstrate that the surface chemistry of the ZrB$_2$ matrix material largely shapes the interface structures (of either Zr-C-Zr or 
B-C-B type) and the nature of the interfacial interaction. The Zr-C-Zr interfaces present strong chemical bonding and their response
to mechanical stress is significantly influenced by graphene corrugation. In contrast B-C-B interfaces, interacting through the relatively 
weak $\pi$-$\pi$ stacking,  show attributes similar to 2D materials heterostructures. Our theoretical results provide insights into the 
interface bonding mechanisms in graphene/ceramic composites, and emphasize the prospect for their design via interface engineering 
enabled by surface contacts.  
\end{abstract}

\section{Introduction} 
Within the last ten years, the use of graphene as nanofiller in ceramic matrix composites (CMCs), the so called GCMC materials, has attracted plenty of research  interest. 
They find application in various industry sectors such as aerospace, automotive, energy \& power, micro-electronics and pharmaceutical.~\cite{Kinloch2018,2013Porwal,SINGH20111178,Walker2011}
 In addition to the excellent mechanical (a tensile strength of 130 GPa and a Young's modulus of 1 TPa), electronic and thermal properties,
 the extremely high specific surface area (2630 m$^2$g$^{-1}$) of graphene provides great capacity for functionalizing and bonding to the surrounding ceramic matrices.~\cite{Tao201709,GAO20171} 
 For instance, the fracture toughness parameter, $K_{IC}$, can be improved by as much as 235~\% for only a 1.5 vol\% addition of graphene in a Si$_3$N$_4$ matrix~\cite{Walker2011}. 
Toughness improvement is also found for zirconium diboride (ZrB$_2$)~\cite{Yadhukulakrishnan2013,AN2016209,An2016}
silicon carbide~\cite{MIRANZO20131665}, tantalum carbide~\cite{NIETO2013338} and alumina~\cite{Liu2016-G}. 
At the same time, the addition of graphene can also suppress the growth of unwanted oxide layers and refine the ceramic 
grains~\cite{Yadhukulakrishnan2013,Liu2013}. Last but not least, the GCMCs developed with hierarchical architectures can simultaneously
improve  the mechanical and functional properties~\cite{Picot2017,AN2016209,Ru2018-Fan}.

Among various ceramic materials that can be benefited from graphene-based nanofillers, ZrB$_2$ classified as ultra-high temperature ceramic 
(UHTC),  is one of the most promising structural ceramics for aerospace propulsion systems~\cite{Scitli2018,Padture2016}. 
It exhibits unique combination of high melting point (T$_m \sim$ 3246 $\degree$C), chemical inertness, effective wear and environment resistance. 
However, the relatively weak fracture toughness and the drop of flexural strength and oxidation resistance at high temperatures~\cite{Dai2017}
are awaiting further improvement. Adding continuous fibers (for enhancing fractural toughness and flexural strength)~\cite{Zoli2018} and 
nano-particles (such as SiC for improving oxidation resistance)~\cite{HU20092724} can partially overcome these issues. 
Very recently, the incorporation of graphene into ZrB$_2$ matrix~\cite{Yadhukulakrishnan2013,AN2016209,An2016} found great
prospect of property enhancement via interfacial impacts.
Although it was reported that the interfacial shear strength can be enhanced by $\sim$ 236\% and the tensile strength by $\sim$ 96\% via coating graphene materials.~\cite{2018Sarker} 
In most cases, the interfacial characteristics is still vague since the characterization techniques of interfaces are generally in an early
development stage~\cite{HATTA20052550,Sheldon2004,kabel2018}.
The difficulty here comes from the nano-size and morphology of interfaces, the presence and variation of defects along  interfaces,
the sophisticated interface alignments during tests, as well as the complexity associated to data deconvolution and deviations from physical models.


To crack the nuts related with interfacial mechanics, the atomistic simulation method offers a valid alternative. 
The technique of molecular dynamics (MD) has been applied to examine NbC/Nb\cite{Salehinia2015}, ZrB$_2$/ZrC\cite{Kayser2018} and 
ZrB$_2$(0001)/graphene interfaces~\cite{An2015}. For instance, the bonding energies of the ZrB$_2$(0001)/graphene interface 
have been predicted by using a universal potential function (a purely diagonal harmonic force field). 
The first-principles calculations were adopted to investigate new mechanical and chemistry phenomena related with interfaces and interphase.~\cite{Shahsavari2018,Shi2017-Zhao,Zhu2015-Mo}  
The large enhancement of strength, ductivlity  and resilience of nano-layered h-BN/silcates was demonstrated by Shahsavarito using the horizontally stacked nanolaminate model.  

Here, we exhibited that the interfacial strength of graphene-reinforced ZrB$_2$ nanocompoiste can be largely engineered for more than one order
of magnitude by tuning the contact surfaces. This origins in the variation of interfacial bonding mechanism, covalent bonded Zr-C-Zr
interfaces or B-C-B interfaces with weak $\pi$-$\pi$ stacking. Also, the corrugation of graphene can further modify the deformation behavior
of Zr-C-Zr interfaces, which is different for the processes of interfacial opening and sliding. In comparison, B-C-B interfaces are
not that sensitive to the rippling of graphene. We highlighted that the enhancement of interfacial properties of graphene/ZrB$_2$ nanocomposite
(a typical example of GCMC) is viable by tuning the chemical environment (leading to a rich variety of Zr-, B- and mix-terminated
surfaces) and the interfacial strains (resulting in various extent of graphene ruga).  This kind of GCMC materials when properly tailored can
be a multifunctional nanocomposites with superior characteristics such as mechanical and electrical properties, thermal and radiation tolerance.

\section{Methodology} 
In this work, we build twelve interface models presenting the tri-layer structures ZrB$_2$/Gr /ZrB$_2$ (Gr = graphene), as schematically 
shown in Fig.~\ref{fig1:config}(a). Since the direct description of misfit dislocations is beyond the current calculation capability of 
DFT (supercells with too many atoms are needed), we adopt here the commensurate interface model, in which the ZrB$_2$ 
slab and the graphene monolayer are constrained to have a common lateral lattice parameter. In brief, three types of interface 
supercells (I, II and III) are constructed by exposing  graphene to the three most stable surfaces of the ZrB$_2$ matrix. These
are respectively the Zr- and B-terminated (0001)  and the Zr-terminated (10$\bar{1}$0) surfaces. The two ends of 
such hybrid structures are separated by a vacuum region of 16~\AA, in order to prevent the fictitious interaction between the
periodic replicas. The two surfaces facing the vacuum regions are all Zr terminated so to reduce the possible effects arising
from surface dipoles and ruffling.  The structural details of these interface models will be elaborated in section~\ref{3.1}.

First-principles calculations are performed within the DFT framework using the plane-wave basis projector augmented wave 
method~\cite{Blochl1994} as implemented in the VASP code~\cite{Kresse1999}. The generalised gradient approximation (GGA) 
parameterised by Perdew, Burke and Ernzerhof (PBE)~\cite{Perdew1996} provides the exchange-correlation energy and 
potential. In addition, damped van der Waals (vdW) corrections (DFT-D2)~\cite{Grimme.2006} are included to account 
for dispersion interactions. The reliability of the PBE+D2 method in describing transition metal di-borides and graphite has been 
established before~\cite{Zhang2018,Chen2013c}.

The Brillouin zone of our interface models are sampled by using the Monkhorst-Pack \textit{k}-point method, with the following
\textit{k}-meshes, 16$\times$16$\times$1, 14$\times$14$\times$1 and 12$\times$7$\times$1, respectively for supercells I, II 
and III. The plane-wave kinetic energy cutoff is set to 500~eV.  These convergence parameters have all been tested to ensure 
an energy convergence of 1~meV/atom.

\section{Results} 
\subsection{Interface configurations}\label{3.1}

The ZrB$_2$/Gr/ZrB$_2$ hybrid structures have been constructed with two possible orientations, namely (0001)ZrB$_2$//(0001)Gr 
and (10$\bar{1}$0)ZrB$_2$//(0001)Gr. When constructing the supercells, we have considered both Zr- and B-terminated ZrB$_2$ 
facing the graphene layer for the (0001)ZrB$_2$ surface, while only Zr-termination is investigated for (10$\bar{1}$0)ZrB$_2$. 
As such the interface supercells bind the (0001)$_\mathrm{Zr}$,  (0001)$_\mathrm{B}$ and (10$\bar{1}$0)$_\mathrm{Zr}$ surfaces 
of ZrB$_2$ to graphene (the subscript here indicates the chemical termination of a given surface). These surfaces have been identified 
as having high thermodynamic stability in previous DFT calculations~\cite{Zhang2018} and have been frequently found in experiments~\cite{TENGDELIUS201671,doi:10.1116/1.4916565}. 
They are the most likely ones to be exposed to the bonding with 
graphene nanosheets. 
In addition, the surfaces of (11$\bar{2}$2)$_\mathrm{Zr+B}$ with mixed termination of Zr and B and  (11$\bar{2}$3)$_\mathrm{Zr}$ show a
relatively high stability when the chemical environment is properly adjusted~\cite{Zhang2018}.

\begin{figure}
\centering
\includegraphics[width=6.5in]{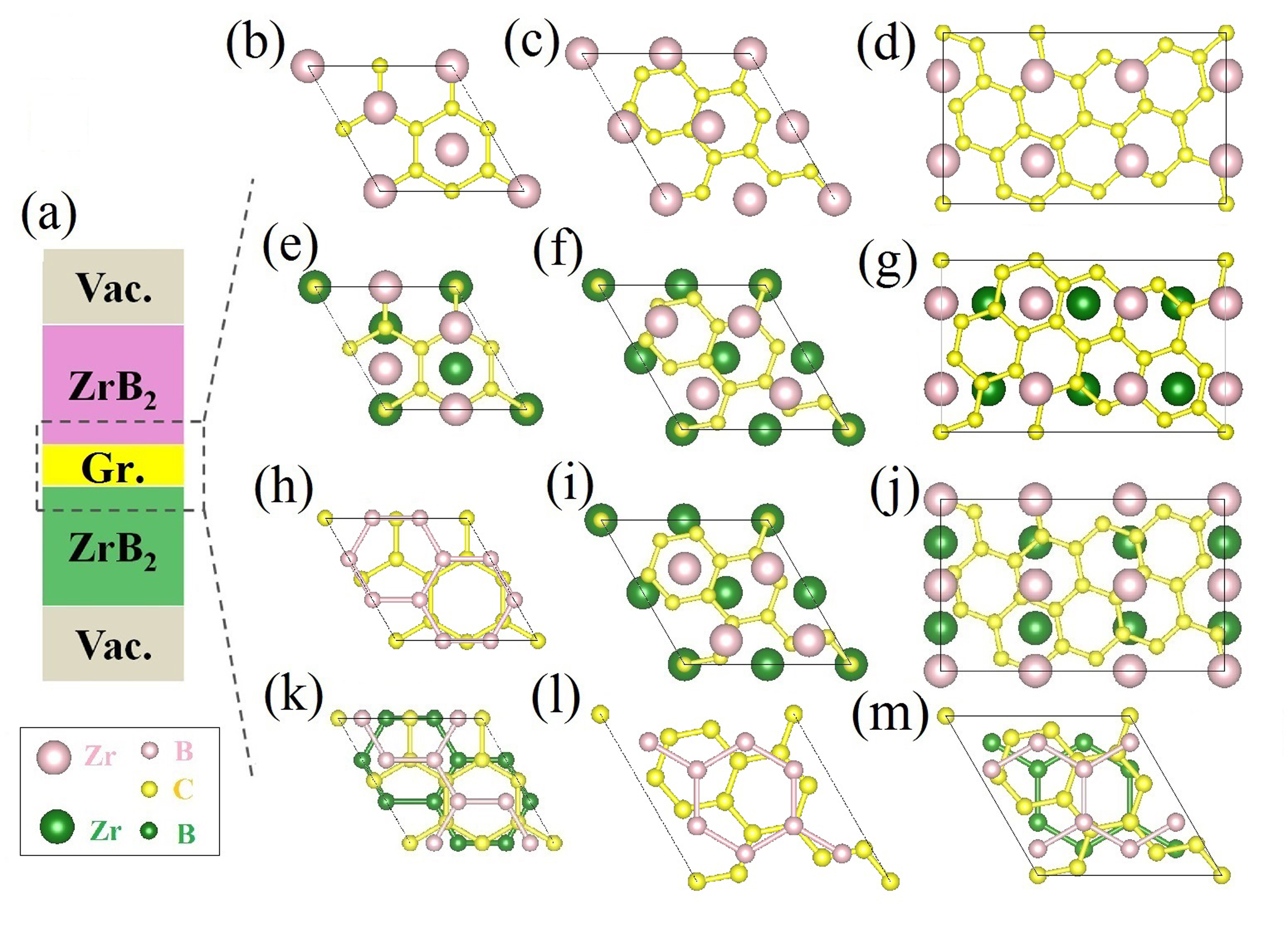}
\caption{(color online) Summary of the interface supercells constructed in this work (a) a schematic drawing of the tri-layer structure 
used in our simulations: the top(bottom) ZrB$_2$ slabs are highlighted in pink (green); while the graphene (Gr) layer between them 
is in yellow. The top view of interface regions enclosed in the dashed box are illustrated in panels (b) through (m):
(b) I$_\mathrm{Zr-C}^\mathrm{AA}$, 
(c) II$_\mathrm{Zr-C}^\mathrm{AA}$,  
(d) III$_\mathrm{Zr-C}^\mathrm{AA}$, 
(e) I$_\mathrm{Zr-C}^\mathrm{AB}$, 
(f) II$_\mathrm{Zr-C}^\mathrm{AB}$,
(g) III$_\mathrm{Zr-C}^\mathrm{AB}$, 
(h) I$_\mathrm{B-C}^\mathrm{AA}$, 
(i) II$_\mathrm{Zr-C}^\mathrm{AC}$, 
(j) III$_\mathrm{Zr-C}^\mathrm{AB'}$,  
(k) I$_\mathrm{B-C}^\mathrm{AB}$, 
(l) II$_\mathrm{B-C}^\mathrm{AA}$,  
(m) II$_\mathrm{B-C}^\mathrm{AB}$. The small yellow spheres are the graphene C atoms. The small and large pink (green) spheres 
are for the B and Zr atoms, respectively, located on the top (bottom) ZrB$_2$ slab. See the text for the convention used to define 
various interfaces. Note that the models (g) and (j) are both AB type, but present an inequivalent atomic arrangement.}
\label{fig1:config}
\end{figure}

Graphene and ZrB$_2$ slabs are joined to form a interface model by using the coincidence lattice method~\cite{doi:10.1021/acs.jpcc.6b01496}. 
In a nutshell this consists in rotating and straining graphene and ZrB$_2$ supercells so to obtain one supercell with common lattice
vectors and little lattice mismatch that, as the same time, with a limited number of atoms. Such exercise has returned us 
three optimum supercells, where the ZrB$_2$ slabs are as follows:  I) $\sqrt{3}\times\sqrt{3}$  (0001); II) $2\times2$ (0001) and 
III) $2\times3$ (10$\bar{1}$0). These define three different interface models, labelled as I, II and III [see figure~\ref{fig1:config} 
panels (b), (c) and (d)], with a lattice mismatch of $\Delta a=10$\%, $\Delta a=2$\% and $\Delta a=3$\% ($\Delta b=6$\%), 
respectively ($a$ and $b$ are the in-plane lattice parameters). A second characteristic defining the interfaces is the stacking
order at the ZrB$_2$ ends. 
We denote as AA the situation where the two ZrB$_2$ blocks neighbouring graphene are symmetric, so that Zr atoms
 of the upper slab is on top of the same kind of atoms on the lower one [see Figs.~\ref{fig1:config}(b),
\ref{fig1:config}(c), \ref{fig1:config}(d), \ref{fig1:config}(h) and \ref{fig1:config}(l)]. In contrast, the stacking is called AB 
type when the Zr atoms of top ZrB$_2$ slab are in a bridge position with respect to those at the bottom [see 
\ref{fig1:config}(e), \ref{fig1:config}(f), \ref{fig1:config}(g), \ref{fig1:config}(j), \ref{fig1:config}(k) and \ref{fig1:config}(m)].
Finally AC stacking denotes the situation where the atoms in top ZrB$_2$ slab are in a hollow position [see 
\ref{fig1:config}(c)].
The thicknesses of the various interfaces are set between 17 and 26 atomic layers (see the thickness test in the SI of 
Ref.~\cite{Zhang2018}). As already mentioned the two ends of such hybrid structures are separated by a vacuum region of 16~\AA, 
in order to prevent the fictitious interaction between the periodic replicas. It is known that the Zr-terminated surfaces have low 
surface strains~\cite{Zhang2018}. Therefore, the two ZrB$_2$ surfaces facing the vacuum regions are all Zr terminated so to 
reduce the possible effects arising from surface dipoles and ruffling. 

Taking all this into consideration our notation used to describe the hybrid structures is based on 1) the supercell types, 2) the atomic 
species facing at the interface, and 3) the stacking sequence. For example, II$_\mathrm{Zr-C}^\mathrm{AA}$, 
II$_\mathrm{Zr-C}^\mathrm{AB}$ and II$_\mathrm{Zr-C}^\mathrm{AC}$ describe the interface models having type II supercell, 
Zr-C-Zr facing species and the stacking sequences AA, AB and AC, respectively. The top view of their interface regions are illustrated 
in the panels (c), (f) and (i) of Fig.~\ref{fig1:config}, which contains the same plots for all the interfaces investigated. To be more specific, 
the atomic structures of our interface models have the following characteristics:

\begin{itemize}

\item I$_\mathrm{Zr-C}^\mathrm{AA}$ and I$_\mathrm{Zr-C}^\mathrm{AB}$ [86 atoms with chemical composition 
(Zr$_3$B$_6$)$_8$C$_8$Zr$_6$] have Zr-C-Zr bonding across both interfaces. They have the same in-plane arrangement 
as the Ni(111)/graphene interface shown in Fig.~1(b) of Ref.~\cite{Giovannetti2008}. The supercells are made of two fragments 
of the \space $\sqrt[]{3}$ $\times$ $\sqrt[]{3}$ (0001) ZrB$_2$ slab, sandwiching a graphene monolayer rotated by 30\degree \space 
with respect to the ZrB$_2$ borophene plane [see Figs.~\ref{fig1:config}(b) and \ref{fig1:config}(e)];

\item I$_\mathrm{B-C}^\mathrm{AA}$ and I$_\mathrm{B-C}^\mathrm{AB}$ [89 atoms with composition (Zr$_3$B$_6$)$_9$C$_8$], 
displayed in Figs.~\ref{fig1:config}(h) and \ref{fig1:config}(k), are similar to  I$_\mathrm{Zr-C}^\mathrm{AA}$ and I$_\mathrm{Zr-C}^\mathrm{AB}$, 
except that the interfaces are B-C-B type;

\item II$_\mathrm{Zr-C}^\mathrm{AA}$, II$_\mathrm{Zr-C}^\mathrm{AB}$ and II$_\mathrm{Zr-C}^\mathrm{AC}$ [118 atoms with composition (Zr$_4$B$_{8}$)$_8$C$_{14}$Zr$_{8}$], shown in Figs.~\ref{fig1:config}(c), \ref{fig1:config}(f) and \ref{fig1:config}(i), are built from two 
blocks of the 2 $\times$ 2 (0001) ZrB$_2$ slab and one graphene monolayer. Here, the graphene is rotated by 19.1\degree \space with 
respect to the ZrB$_2$ borophene plane. They have the Zr-C-Zr [0001]$_\mathrm{ZrB_2}$ stack similarly to I$_\mathrm{Zr-C}^\mathrm{AA}$ 
and I$_\mathrm{Zr-C}^\mathrm{AB}$.  However, the Zr atoms are misaligned with the C atoms in graphene, resulting in a smaller lattice 
mismatch (2\% against 10\%);

\item II$_\mathrm{B-C}^\mathrm{AA}$ and II$_\mathrm{B-C}^\mathrm{AB}$ [110 atoms with composition (Zr$_4$B$_{8}$)$_8$C$_{14}$] 
are similar to II$_\mathrm{Zr-C}^\mathrm{AA}$ and II$_\mathrm{Zr-C}^\mathrm{AB}$, except that the interface structure is B-C-B 
type [see Figs.~\ref{fig1:config}(l) and \ref{fig1:config}(m)];

\item III$_\mathrm{Zr-C}^\mathrm{AA}$, III$_\mathrm{Zr-C}^\mathrm{AB}$ and III$_\mathrm{Zr-C}^\mathrm{AB'}$ [148 atoms with 
composition (Zr$_6$B$_{12}$)$_6$C$_{28}$Zr$_{12}$], shown in Figs.~\ref{fig1:config}(d), \ref{fig1:config}(g) and \ref{fig1:config}(j), 
are constructed from the 2 $\times$ 3 (10$\bar{1}$0) surface slab and one graphene monolayer. They have the Zr-C-Zr [10$\bar{1}$0]$_\mathrm{ZrB_2}$ termination, and the Zr atoms are misaligned with respect to the C atoms in graphene. 

\end{itemize}

Finally, the ZrB$_2$/graphene interlayer distance, $d_i$, and the common lattice parameters ($a_i$ and $b_i$) are optimized by 
searching for the lowest energy points of the $E(a_i, b_i, d_i)$ potential energy surface (see section~\ref{3.2}). The so-calculated 
parameters are tabulated in Table S1 of the supplementary information (SI), together with the in-plane strains. At such optimized 
lattice parameters the ionic positions are fully relaxed into their ground state by using the quasi-Newton algorithm to relieve the 
residue stresses. 

\subsection{Interface Energetics} 
%
%
\subsubsection{Interfacial adhesion energy ($E_{ad}$)}
\label{3.2}

The adhesion energy, $E_\mathrm{ad}$, is defined as the energy per unit area released when forming a multi-layered structure 
from the isolated surface slabs, namely
\begin{align}\label{Eb1}
E_\mathrm{ad} = \frac{E_\mathrm{tot}(a_i)-E_{\mathrm{top}}(a_0)-E_{\mathrm{bot}}(a_0)-E_{\mathrm{Gr}}(l_\mathrm{C-C})}{2A}\:.
\end{align}
\noindent Here, E$_\mathrm{tot}$(a$_i$), E$_{\mathrm{top}}$(a$_0$),  E$_{\mathrm{bot}}$(a$_0$) and 
E$_{\mathrm{Gr}}$($l_\mathrm{C-C}$) are, respectively, the total energy of the hybrid structure, that of 
the top and bottom ZrB$_2$ slab and of the graphene monolayer. Note that the reference ZrB$_2$ slabs 
have an in-plane lattice parameter, $a_0$, corresponding to that of their strain-free surface configuration. 
In contrast the reference graphene monolayer has a C-C bond length of $l_\mathrm{C-C}=1.42$~\AA.
As a consequence of this choice $E_\mathrm{ad}$ for the various hybrid structures is computed relatively
to the same reference states, namely the strain-free surface slabs. An alternative choice is to take the bulk 
structures as reference, a choice that will include the surface formation energy into the definition of $E_\mathrm{ad}$.
Finally, $A$ is the interface area and the pre-factor 2 takes into account the fact that our hybrid structures have 
two interfaces. 

\begin{figure}
\centering
\includegraphics[width=6.2in]{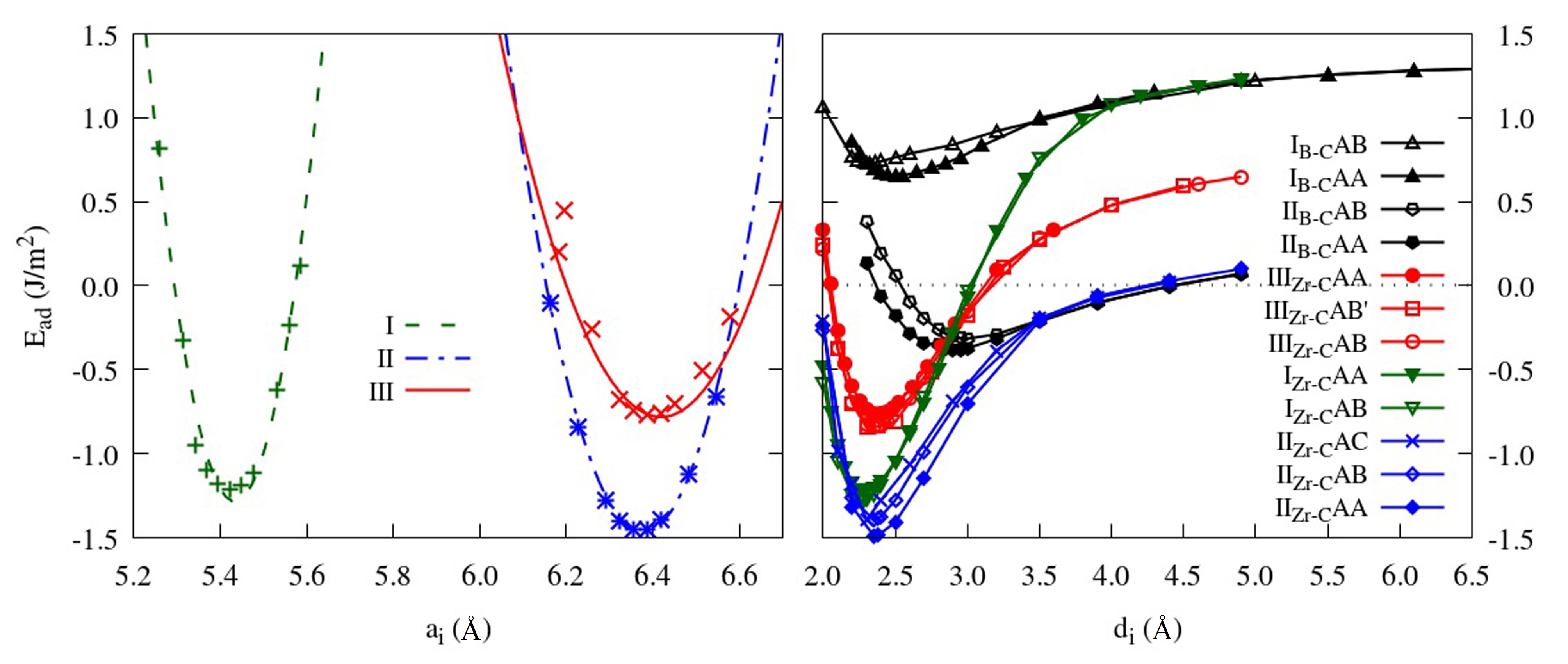}
\caption{(color online) The variation of the adhesion energy, $E_\mathrm{ad}$ (in J/m$^2$), as a function of the (a) in-plane lattice 
parameter, $a_i$ (\AA), and (b) the interlayer distance, $d_i$ (\AA). Results are plotted for the three supercell types and for the different 
interface models of Fig.~1.}
\label{fig2:Ead}
\end{figure}

The calculated $E_\mathrm{ad}(a_i; d_i)$ curves are presented in Fig.~\ref{fig2:Ead}(a) for all the three kinds of interface 
supercells I, II and III, while the interlayer distances, $d_i$, are kept fixed at 2.5 \AA.  A parabolic behaviour is observed 
in all cases, similarly to what recently reported for the interface between graphene and Ti$_2$C MXene \cite{Paul2017}. 
The in-plane lattice parameters of the various interfaces are thus determined from the minima of Fig.~\ref{fig2:Ead}(a). 
Then the optimal interlayer distances, $d_0$, are computed by looking at the minima of the $E_\mathrm{ad}(d_i; a_i)$ curves
taken at the optimized $a_i$, see Fig.~\ref{fig2:Ead}(b). 

Several comments can be made by looking at Fig.~\ref{fig2:Ead}(b). Firstly, we note that the overall $E_\mathrm{ad}$ curves move 
to a lower energy as we go across the series I$_\mathrm{B-C}$, II$_\mathrm{B-C}$, III$_\mathrm{Zr-C}$, I$_\mathrm{Zr-C}$ and 
II$_\mathrm{Zr-C}$. Their thermal stability therefore has to be ranked in the reverse order. Secondly, it is quite clear that all the Zr-C 
interfaces have a deeper $E_\mathrm{ad}$ potential well than those with B-C bonding, suggesting that the Zr-C interfaces are energetically 
more favorable than the B-C ones. Thirdly, we find that the exact stacking order has little effect on the $E_\mathrm{ad}$ curves, in
particular on their energy minimum, indicating that the local bonding environment plays only a minor role in the interface stability.
Finally, one has to note that all the $E_\mathrm{ad}$ curves have a long distance tail that asymptotically converges to a positive 
$E_\mathrm{ad}$ value. In particular we have all type I curves converging to $U_\mathrm{I} = 1.288$~J/m$^2$, the type II to 
$U_\mathrm{II} = 0.261$~J/m$^2$ and the type III to $U_\mathrm{III} = 0.838$~J/m$^2$. Such asymptotic values, $U_i$,
correspond to the misfit strain energies introduced by imposing a common in-plane lattice parameter. Thus the $U_i$'s are 
proportional to the misfit strains associated to the various interface models. In fact the strain energies are ranked in the order 
$U_\mathrm{I}>U_\mathrm{III}>U_\mathrm{II}$, which is the same order of the in-plane strains.  

Although it is too computationally expensive to include misfit dislocations in our DFT calculations, because of the large supercells
required, the misfit strain can be effectively released by considering configurations where graphene presents vertical corrugation, 
an intrinsic feature of graphene flakes~\cite{Meyer2007}. After full structural relaxation, the Zr-C interfaces with AB and AC 
stacking orders exhibit a more pronounced graphene corrugation than that corresponding to the AA stacking. In contrast, for 
B-C interfaces more pronounced graphene rippling is present for the AA order. The underlying mechanisms leading to these 
structural differences will be analyzed later when discussing the electronic structure of the interfaces.

\subsubsection{Interfacial binding energy}
\label{3.2.2}
As discussed before our interface supercells are constructed using the commensurate model so that the heterogeneous layers have common 
in-plane lattice parameters, a fact that introduces misfit strain. As a consequence, contributions to $E_\mathrm{ad}$ originating 
from the misfit strain energy, $U_i$, add to those coming from the formation of the chemical bond at the interface, $E_\mathrm{b}$. 
In order to decouple the two contributions, we assume that $E_\mathrm{ad}$ can be written as 
\begin{align}\label{Ead-M}
E_\mathrm{ad} = - E_\mathrm{b} + U_i   \:. 
\end{align}
The misfit energy $U_i$ can be approximated by the following expression
\begin{align}\label{Ui}
U_i (a_i) =  \frac{[E_\mathrm{top}(a_i)-E_{\mathrm{top}}(a_0)]+[E_{\mathrm{bot}}(a_i)-E_{\mathrm{bot}}(a_0)]+[E_{\mathrm{Gr}}(a_i)-E_{\mathrm{Gr}}(l_\mathrm{C-C})]}{2A}\   \:,
\end{align}
where $E_\mathrm{top}(a_i)$, $E_{\mathrm{bot}}(a_i)$ and $E_{\mathrm{Gr}}(a_i)$ are, respectively, the total energy of 
the top and bottom ZrB$_2$ slab and that of the graphene monolayer. The other terms are the same as mentioned in Eq. (1). Hence one has
\begin{align}\label{Eb}
-E_\mathrm{b}(d_i; a_i) = E_\mathrm{ad} (d_i; a_i) -  U_i (a_i) = \frac{E_\mathrm{tot}(d_i;a_i)-E_{\mathrm{top}}(a_i)-E_{\mathrm{bot}}(a_i)-E_{\mathrm{Gr}}(a_i)}{2A}
 \:. 
\end{align}

\begin{figure}
\centering
\includegraphics[width=4.2in]{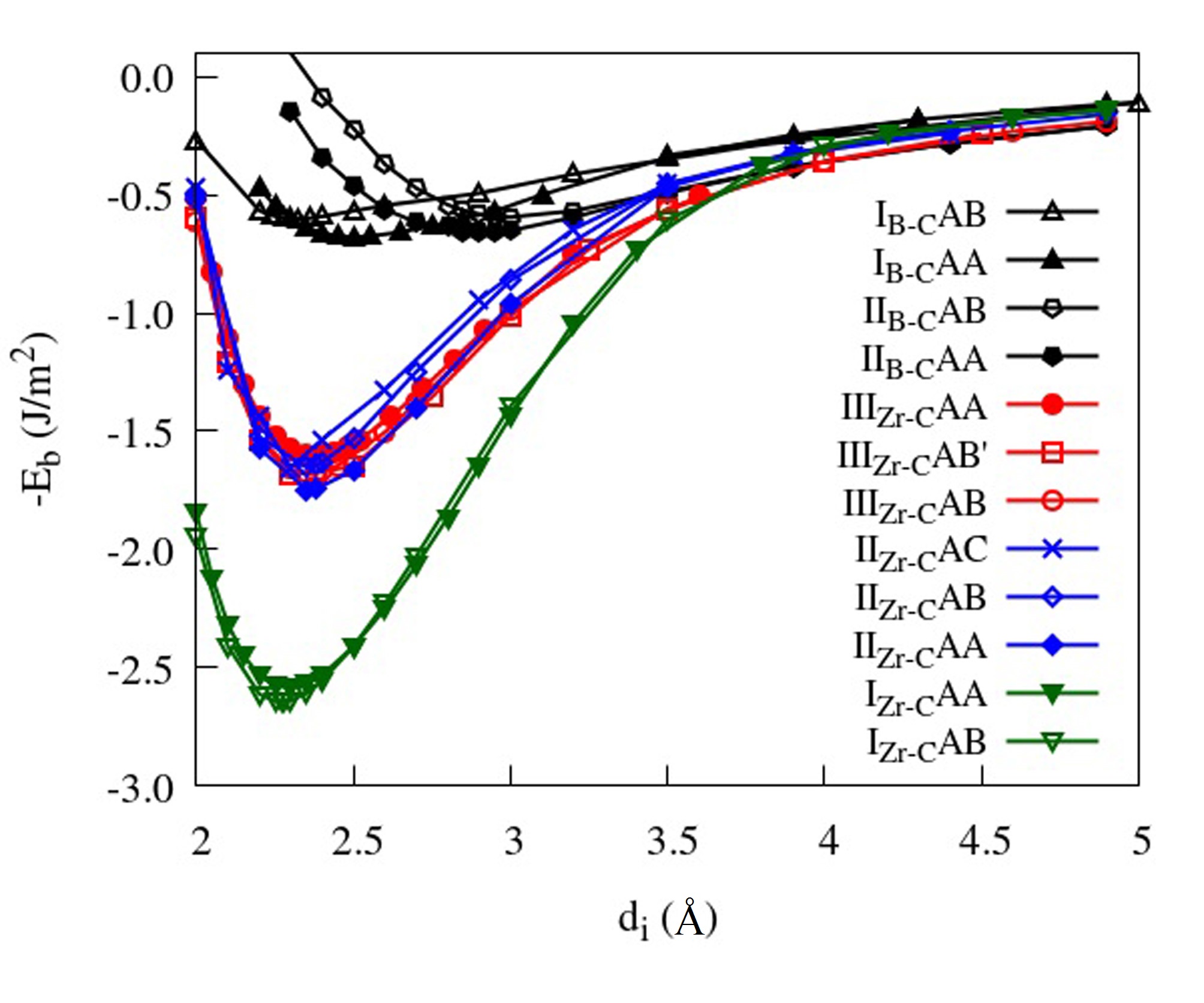}
\caption{(color online) Interfacial binding energy, -$E_b$, (J/m$^2$) against the interlayer separation distance, $d_i$, (\AA).}
\label{fig3:Eb}
\end{figure}

The binding energy curves $E_\mathrm{b}(d_i; a_i)$ are then presented in Fig.~\ref{fig3:Eb}. In comparison of
$E_\mathrm{ad}(d_i)$, they all asymptotically converge to zero. Interestingly all the $E_\mathrm{b}(d_i)$ curves seem to cluster 
into three main groups: (i) the I$_\mathrm{Zr-C}$ interface has the deepest $E_\mathrm{b}$ well ($\sim2.6$~J/m$^2$), indicating
strong interfacial interaction; (ii) the interfaces II$_\mathrm{Zr-C}$ and III$_\mathrm{Zr-C}$ have an intermediate $E_\mathrm{b}$ 
minimum at around 1.7~J/m$^2$, which is 35\% lower than that of I$_\mathrm{Zr-C}$, suggesting a moderate interfacial interaction; 
(iii) all the interfaces of B-C type show a very shallow $E_\mathrm{b}$ profile, with a minimum at around 0.5~J/m$^2$. This latter value 
is close to that calculated for graphite (0.4 J/m$^2$). 

It is worthy to mention that $E_\mathrm{ad}$ and $E_\mathrm{b}$ for the B-C interfaces are about one order of magnitude smaller than 
those of the Zr-C interfaces. Such differences are related to the interfacial bonding mechanism. Our guess is that graphene/borophene 
layers in the B-C interface are coupled by weak physical adsorption, while the Zr-C interfaces are bonded by strong chemical interaction, 
similar to that at play in metal/graphene contacts~\cite{Giovannetti2008}. These hypotheses will be investigated further in the electronic 
structures section~\ref{4.1}.
In concluding this section we remark our main finding, namely that the interface adhesion energy, $E_\mathrm{ad}$, is strongly 
affected by the chemical species facing each other across the interface. This, however, also includes the misfit strain energy, $U_i$. 
When $U_i$ is subtracted from $E_\mathrm{ad}$ we are left with the interface binding energy, $E_\mathrm{b}$, which ranks 
the stability of our interface models as I$_\mathrm{Zr-C}>\mathrm{II/III}_\mathrm{Zr-C}>\mathrm{I/II}_\mathrm{B-C}$. 
It is worthy to mention that, besides the thermodynamically stable Zr-C interfaces, also the B-C ones may be relevant for
composites, since B-terminated ZrB$_2$~\cite{0953-8984-20-26-265006} can form in B-rich growth conditions. %

\subsection{Interfacial cleavage}
The interfacial fracture behavior is investigated by computing the traction-displacement curves, 
which are obtained from the derivative of the total energy changes upon displacing two materials 
adjacent to the interface~\cite{Jiang2010}. Calculations are performed for two different displacement modes, namely
opening (mode I, the two sides of the interface are displaced orthogonally to the interface plane) and 
sliding (mode II, the two sides of the interface are displaced along the interface plane). In this section 
we will present results for the loading mode I, for which we have adopted the same cleavage model of 
Lazar and Podloucky~\cite{Lazar2008} for bulk materials, while the displacement mode II (sliding) will be discussed 
in section~\ref{3.4}. In the fracture mode I the tri-layer structure is separated by introducing an initial 
displacement of length $z$ between the top ZrB$_2$ slab and the graphene layer, so that a pre-existing 
crack is introduced at one side of the interface. Thus, the inter-layer distances of the graphene layer with 
the bottom and top ZrB$_2$ slabs are, respectively $d_0$ and $d_0+z$, where $d_0$ is the equilibrium interlayer
distance calculated before. The corresponding energy change per unit area defines the cleavage energy, 
$E_\mathrm{c}$, which writes
\begin{align}\label{dE}
E_\mathrm{c}(z) = \frac{E_\mathrm{tot}(d_0+z) - E_\mathrm{tot} (d_0)}{A} \:.
\end{align}
In Eq.~(\ref{dE}) $E_\mathrm{tot}(d_0)$ and $E_\mathrm{tot}(d_0+z)$ are the total energies of the pristine 
hybrid structure and of the cleaved one, respectively. 

We have then fitted the $E_\mathrm{c}(z)$ curves with a Morse function
\begin{align}\label{Ec}
E_\mathrm{c}(z)=W_\mathrm{sep}\times[1-exp^{-a^\prime z}]^2 \:,
\end{align}
where $W_\mathrm{sep}$ is the work of separation. Here the parameter $a^\prime$ determines the width of the 
$E_\mathrm{c}(z)$ curve, while $2 W_\mathrm{sep} \times (a^\prime)^2$ controls its curvature at $z = 0$. 
The traction curve, $\sigma(z)$, is calculated as the first derivative of $E_\mathrm{c}(z)$ with respect to $z$
\begin{align}\label{bc}
\sigma(z)=20a^\prime W_\mathrm{sep}\times[exp^{-a^\prime z}-e^{-2a^\prime z}] \:,
\end{align}
where $\sigma(z)$ is in GPa.
Then, the interfacial cleavage strength, $\sigma_\mathrm{c}$, and the critical crack length, $\delta_\mathrm{c}$, are 
defined as the values of $\sigma$ and $z$ at the maximum of the $\sigma(z)$ curve. 
The general behaviour of $\sigma(z)$ is rather simple. As the pre-opening crack grows ($z$ gets larger), $E_\mathrm{c}$ continuously increases 
until it reaches its maximum value, $W_\mathrm{sep}$. Thereafter a crack between free (non-interacting) surfaces is 
formed. The final separation, $\delta_\mathrm{f}$, thus can be written as
\begin{align}\label{df}
\delta_\mathrm{f} = \frac{2 \Delta W_\mathrm{sep}}{\sigma_\mathrm{c}} \:. 
\end{align}

\begin{figure}
\centering
\includegraphics[width=5.2in]{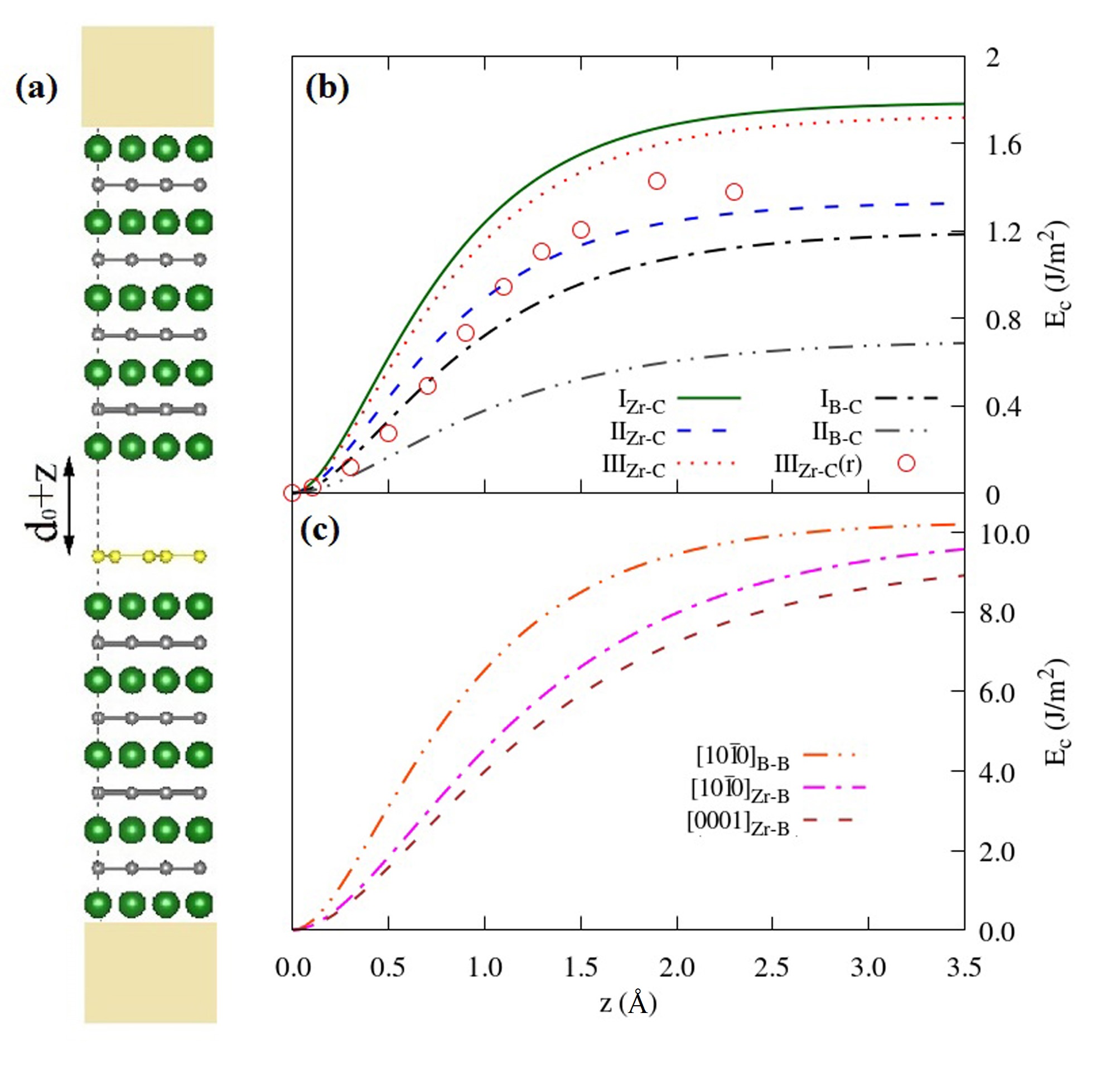}
\caption{(color online) The cleavage of ZrB$_2$/graphene composite. (a) A schematic diagram of the cleavage mode obtained 
by adding a pre-opening crack of length $z$ to the equilibrium tri-layer structures (colour code: green = Zr, grey = B, yellow = C). 
In panel (b) we present the energy separation curves, $E_\mathrm{c}(z)$, for rigid cleaving (no relaxation) between Zr and C, and 
between B and C for various interface models. In panel (c) the same quantity is plotted for cleaving ZrB$_2$ alone between
two B planes and between Zr and B planes along [10$\bar{1}$0] and  [0001] directions. In panel (b) we also report calculations 
where atomic positions are relaxed. This is labelled as III$_\mathrm{Zr-C}$(r) and the curve must be compared with the one for the same 
cleavage but obtained without relaxation, III$_\mathrm{Zr-C}$. }
\label{fig5:bonding}
\end{figure}

\begin{figure}
\centering
\includegraphics[width=4.2in]{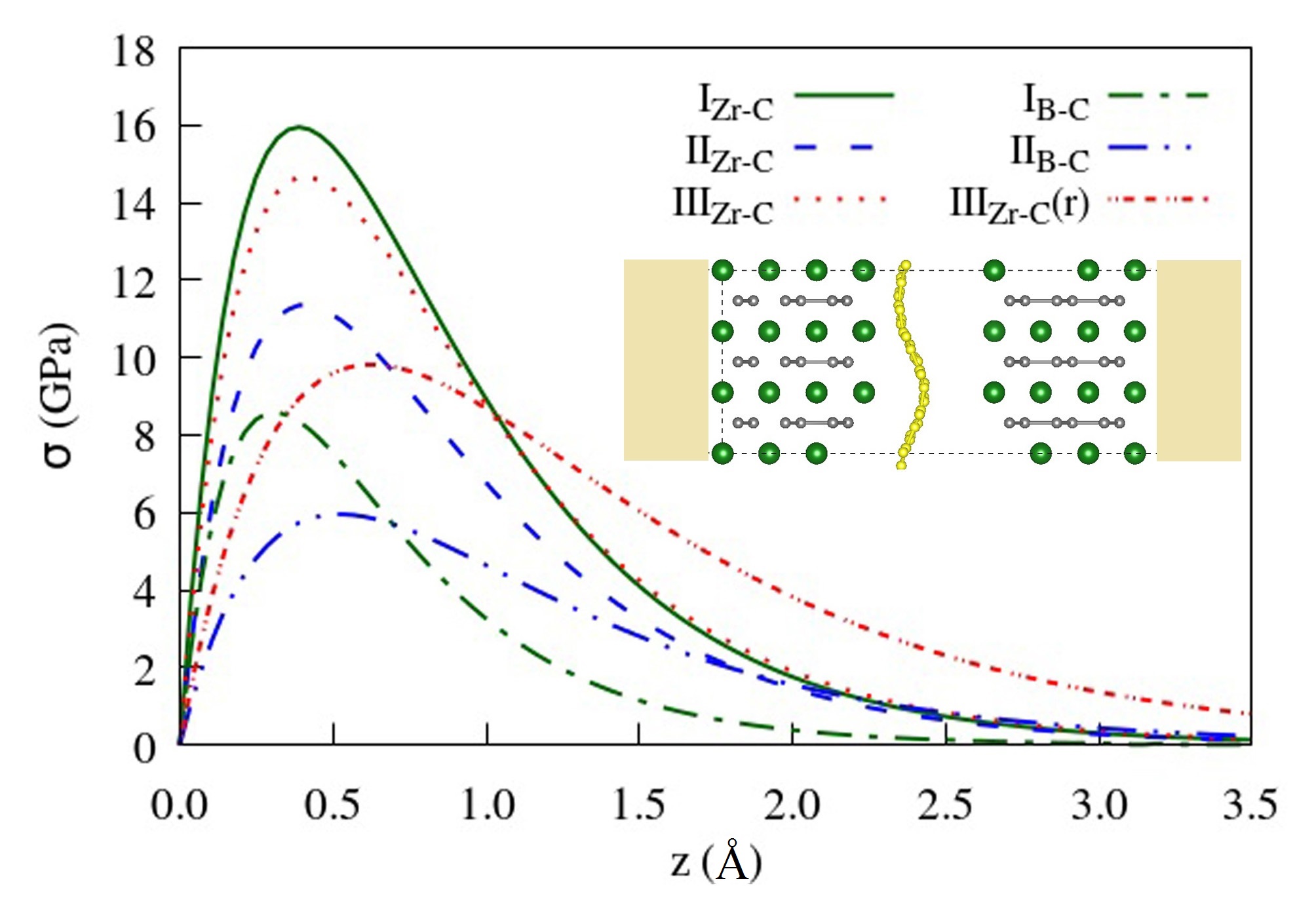}
\caption{(color online) Traction curves for the cleavage between Zr and C of the interface models I, II and III, and for 
the cleavage between B and C of the interface supercells I and II. A relaxation case with a rugged fracture graphene 
surface is labeled as III$_\mathrm{Zr-C}$(r), and the atomic structure next to the fracture surface is presented as a 
ball-and-stick plot in the inset (colour code: green = Zr, grey = B, yellow = C). }
\label{fig6:cleavage}
\end{figure}

The typical mode of cleavage used in the calculation is presented in Fig.~\ref{fig5:bonding}(a), our calculated cleavage 
energies are shown in Fig.~\ref{fig5:bonding}(b) for cleaving across the Zr-C and B-C interfaces, and in Fig.~\ref{fig5:bonding}(c) 
for cleaving Zr-B and B-B atomic layers in the ZrB$_2$ matrix, while the corresponding traction-separation curves are 
displayed in Fig.~\ref{fig6:cleavage}. Note that three kinds of structural relaxation strategies are considered when calculating
the relevant cleavage quantities, and these are explained in the SI. We start our analysis with the brittle cleavage due to a 
sharp fracture surface and then we move our attention to the effects of structural relaxation and of the graphene layer corrugation. 

Among all the cleavage modes, cleaving B-B bonds along [10$\bar{1}$0] in bulk ZrB$_2$ has the highest $W_\mathrm{sep}$ 
value of 10.29~J/m$^2$, while cleaving between Zr and B both along [10$\bar{1}$0] and [0001] has rather similar 
$W_\mathrm{sep}$ ($\sim$9.50~J/m$^2$). In comparison, cleaving across the interfaces with graphene requires
works of separation one order of magnitude smaller (values in the range 0.70-1.79~J/m$^2$). This confirms that 
the ZrB$_2$/graphene interfaces as the weak parts can deflect cracks during the fracture of composites. Such feature 
allows the energy to be released at the ZrB$_2$/graphene interface, so that the structural integrity of the ZrB$_2$/graphene  composite can be preserved.

In more detail, both $W_\mathrm{sep}$ and $\sigma_\mathrm{c}$ systematically decrease when going through the 
interfaces I$_\mathrm{Zr-C}$, III$_\mathrm{Zr-C}$, II$_\mathrm{Zr-C}$, I$_\mathrm{B-C}$ and II$_\mathrm{B-C}$, namely
they follow the ranking obtained from the binding energy curves of Fig.~\ref{fig3:Eb}. This correspondence was also 
observed in the past when studying interfaces in lamellar TiAl alloys~\cite{KANANI2014154}. Since the crack is initiated by 
breaking the interfacial Zr-C or B-C bonds, it appears that the bonding mechanism across the interfaces is responsible for the 
interface strength. 
In other words, the values of $\sigma_\mathrm{c}$, $W_\mathrm{sep}$ and $E_b$ all reflect the ease of interfacial debonding. 
Although in real composites additional features, such as defects and impurities, may modify the strength of the interfacial interaction and then affect the way of crack  propagation, the ranking calculated here already 
provides a clear map of the mechanical stability of the various components of a graphene/ZrB$_2$ composite.

Several important parameters, such as the work of separation, $W_\mathrm{sep}$, the traction strength, 
$\sigma_\mathrm{c}=\mathrm{max}[\sigma(z)]$, the critical separation, $\delta_\mathrm{c}$, and the final separation, 
$\delta_{f}$, can all be extracted from these curves and are summarized in Table \ref{tab:table1}. These quantities, together 
with the shear parameters presented in Table~\ref{tab:table1}, can then be used as inputs in continuum simulation models of 
cohesive zone to provide a complete understanding of the interface de-bonding of GCMC materials.
It is worth mentioning that the $\sigma_\mathrm{c}$ varies from 15.95 GPa to 4.02 GPa when going from I$_\mathrm{Zr-C}$ to II$_\mathrm{B-C}$.
This provides a chance of tuning the mechanical behavior of interfaces in GCMCs over one order of magnitude.
 
\begin{table}[h!]
\small
\centering
\caption{Mechanical parameters of ZrB$_2$ matrix, graphene and ZrB$_2$/graphene interface. $W_\mathrm{sep}$: work 
of separation, $\sigma_\mathrm{c}$: cleavage strength, $\delta_\mathrm{c}$: critical separation distance, $\delta_\mathrm{f}$ 
final separation distance, $\gamma_\mathrm{us}$: unstable stacking fault energy, $\gamma_\mathrm{sf}$:  stacking fault 
energy, $\tau_\mathrm{c}$: shear strength.}
\begin{tabular}{c|c|c|c|c|c}
\hline
\hline
\multicolumn{2}{c|}{Mode I loading (opening)} &  $W_\mathrm{sep}$ (J/m$^2$) & $\sigma_\mathrm{c}$ (GPa) & $\delta_\mathrm{c}$ (\AA) & $\delta_\mathrm{f}$ (\AA) \\
\hline
ZrB$_2$ & [10$\bar{1}$0]$_\mathrm{B-B}$ & 10.29 & 82.16 & 0.43 & 2.50 \\
        & [10$\bar{1}$0]$_\mathrm{Zr-B}$&  9.97 & 56.06 & 0.62 & 3.56 \\
        & [0001]$_\mathrm{Zr-B}$        &  9.38 & 49.54 & 0.66 & 3.79 \\
\hline
Graphene & [0001]$_\mathrm{C-C}$ &         0.36 &  2.27 & 0.55 &  3.17  \\
\hline
Interfaces & I$_\mathrm{Zr-C}$     & 1.79 & 15.95 & 0.39 & 2.24 \\
         & III$_\mathrm{Zr-C}$    & 1.73 & 14.67 & 0.41 & 2.35 \\
          & II$_\mathrm{Zr-C}$    &  1.33 & 11.34 & 0.41 & 2.35 \\
          & I$_\mathrm{B-C}$      & 1.20 & 8.96 & 0.46 & 2.68 \\
          & II$_\mathrm{B-C}$     &  0.85 &  4.02 & 0.73 & 4.23 \\
\hline
\multicolumn{3}{c|}{Mode II loading (sliding)} & $\gamma_\mathrm{sf}$ (J/m$^2$) & $\gamma_\mathrm{us}$ (J/m$^2$) & $\tau_\mathrm{c}$ (GPa) \\
\hline
ZrB$_2$  & \multicolumn{2}{c|}{$\{0001\}<11\bar{2}0>$}  & - & 3.47  &  49.62 \\
         & \multicolumn{2}{c|}{$\{0001\}<10\bar{1}0>$}  & 3.03 & 3.47 &  24.45  \\
\hline
Graphene & \multicolumn{2}{c|}{armchair}              & - & 0.09  & 1.00  \\
         & \multicolumn{2}{c|}{zigzag}                & - & 0.04  & 0.48  \\
\hline
Interfaces &  \multicolumn{2}{c|}{I$_\mathrm{Zr-C}$  $\{0001\}<11\bar{2}0>$}  & - & 0.57 & 4.55   \\
           &  \multicolumn{2}{c|}{I$_\mathrm{Zr-C}$  $\{0001\}<10\bar{1}0>$}  & 0.05 & 0.57 & 8.37  \\
           &  \multicolumn{2}{c|}{I$_\mathrm{B-C}$  $\{0001\}<11\bar{2}0>$}   & 0.04 & 0.10 & 1.87  \\
           &  \multicolumn{2}{c|}{I$_\mathrm{B-C}$  $\{0001\}<10\bar{1}0>$}   & 0.04 & 0.15 & 3.35  \\
\hline
\end{tabular}
\label{tab:table1}
\end{table}

We have then investigated the effects of the structural relaxation on the energetics of the fracture (a detail discussion is provided 
in the SI). The atomic positions within a region 4.5 \AA~vertical to the fracture surfaces are relaxed in order to minimize the 
internal stresses and the total energy. The results are labelled as III$_\mathrm{Zr-C}$(r) in Figs.~\ref{fig5:bonding}(b) and 
\ref{fig6:cleavage}, while the final geometry is shown in the inset of Fig.~\ref{fig6:cleavage}, where it is evident that relaxation
produces a corrugation of the graphene layer. It is experimentally known that graphene sheets may exhibit large corrugations 
when adsorbed on metal surfaces~\cite{Meyer2007}. This is observed here, since the buckling of graphene can efficiently 
relieve the misfit strains across the interface and provides an energy reduction channel alternative to misfit dislocations. 
As illustrated in Fig.~\ref{fig5:bonding}(b) with the open circles, the corrugation of graphene can lower down $W_\mathrm{sep}$ 
by $\sim$ 0.4 J/m$^2$. At the same time it makes the critical cleavage stress going from 15~GPa to 10~GPa. This demonstrates 
that, in general, graphene buckling has the effect of weakening the interface adhesion. The effect originates from the fact that
the interface, namely the surface of graphene in direct contact with the ZrB$_2$ surface, is partially detached so that the 
contact area is reduced by the corrugation. Note that the rippling period of graphene is constrained by the in-plane dimensions 
of the supercell, and that the III$_\mathrm{Zr-C}^\mathrm{AB}$ supercell (148 atoms) is the largest studied here. One then has 
to expect that for planar structures with a larger in-plane cell, and consequently smaller lattice misfit and internal stress, the 
effect of graphene buckling will be in general less pronounced. 
 
\subsection{Interface sliding}
\label{3.4} 

Interfacial sliding processes are studied in order to extract the traction-separation curves under the loading mode II~\cite{Jiang2010}. 
This is modelled by displacing the top ZrB$_2$ slab along a direction parallel to the interface plane, while monitoring 
the total energy  as a function of the sliding vectors. The sliding profile, $\gamma$, can be defined as the change in 
total energy with respect to the energy of the undistorted structure as a function of the sliding vector, namely
\begin{align}\label{GSFE}
\gamma = (E^\mathrm{sh}_\mathrm{tot} - E^0_\mathrm{tot})/A \:,
\end{align}
where $E^0_\mathrm{tot}$ and $E^\mathrm{sh}_\mathrm{tot}$ are the total energies of the undistorted and of the 
distorted structure, respectively. 

\begin{figure}
\centering
\includegraphics[width=6.5in]{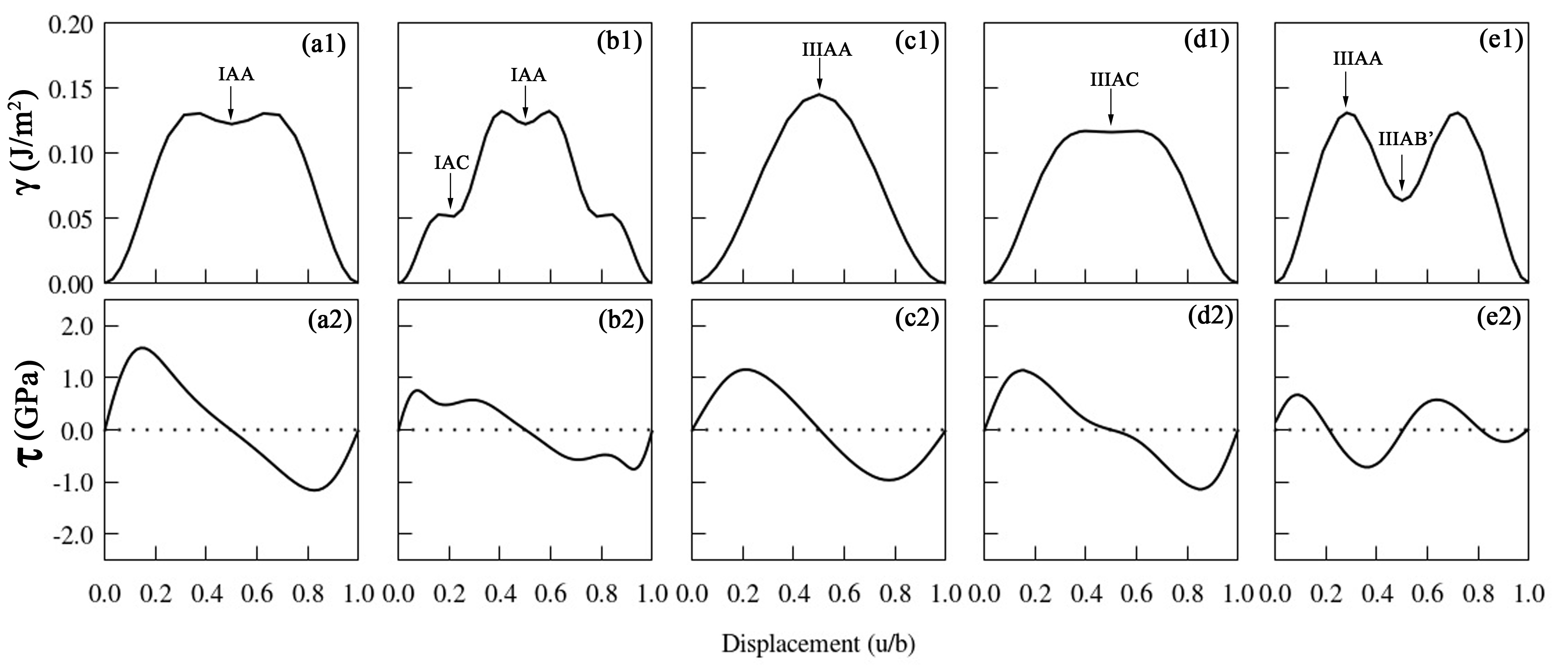}
\caption{Rigid energy landscapes (upper panels), $\gamma$, and corresponding shear stress (lower panels), $\tau$, as a 
function of the displacement. A displacement vector $\vec{u}$ is measured with respect to the relevant Burgers vector, 
$\lvert\vec{b}\rvert$. The Burgers vectors in panels (a) through (e) are, respectively, 
$\vec{b}_1 = \frac{1}{3}\langle\bar{1}2\bar{1}0\rangle_\mathrm{ZrB_2}$, 
$\vec{b}_2 = \langle10\bar{1}0\rangle_\mathrm{ZrB_2}$, 
$\vec{b}_3 = \frac{1}{3}\langle\bar{1}2\bar{1}0\rangle_\mathrm{ZrB_2}$, 
$\vec{b}_4 = \langle0001\rangle_\mathrm{ZrB_2}$ and 
$\vec{b}_5 = \frac{1}{3}\langle11\bar{2}3\rangle_\mathrm{ZrB_2}$.}
\label{fig7:GSFE}
\end{figure}

The {\it rigid} energy landscape is derived by monitoring the energy of the distorted structure without performing 
any structural relaxation. In addition, we have also calculated the effects of full structural relaxation by using the 
nudged elastic band (NEB) method~\cite{Henkelman2000,Henkelman2000a}. In practice we allow atomic relaxation 
both perpendicular to the gliding plane and in-plane and track the minimum energy path (MEP) while keeping the 
Burgers vectors fixed. This fully relaxed calculations have been performed only for the interface models I$_\mathrm{Zr-C}$ 
(86 atoms in the supercell) and I$_\mathrm{B-C}$ (89 atoms). As for the interface types II and III (118 and 148 atoms, 
respectively), only the {\it rigid} sliding profiles have been studied because of the heavy computational costs associated to 
the DFT-based NEB method. The shear stress along a given direction $x$ is then calculated as the slope of the energy 
profile along that direction, namely 
\begin{align}\label{tau}
\tau_x = \frac{d\gamma}{dx} \:,
\end{align}
where the maximum $\tau$ is defined as the interfacial shear strength, $\tau_\mathrm{c}^x=\mathrm{max} \big\{\tau_x\big\}$. 
Five slip systems, namely basal $\langle$a$\rangle$: (0001)$\langle$$\bar{1}$2$\bar{1}$0$\rangle$, 
basal $\langle$b$\rangle$: (0001)$\langle$10$\bar{1}$0$\rangle$, 
prismatic $\langle$a$\rangle$: (10$\bar{1}$0)$\langle$$\bar{1}$2$\bar{1}$0$\rangle$, 
prismatic $\langle$c$\rangle$: (10$\bar{1}$0)$\langle$0001$\rangle$ 
and prismatic $\langle$a+c$\rangle$: (10$\bar{1}$0)$\langle$11$\bar{2}$3$\rangle$, are studied. 
These all concern interfaces of type I and III with [0001]$_\mathrm{ZrB_2}$ and [10$\bar{1}$0]$_\mathrm{ZrB_2}$ 
orientation. The equilibrium geometry of the Zr-C interfaces has an AB stacking sequence, while that of the B-C 
structure is AA type. 
The top panels of Fig.~\ref{fig7:GSFE} displays the energy profiles for {\it rigid} sliding along the Zr-C interfaces. In general 
the energy barriers are observed to be relatively small in magnitude ($<0.15$~J/m$^2$). In addition we find that the 
sliding curves present a number of maxima and minima, marked in the figure by arrows. These correspond to different 
stacking configurations of the supercell (also indicated in the figure), which are connected by the sliding process. As the 
different stacking orders have a rather similar binding energy (see Section~\ref{3.2.2}), we expect the sliding energy profile 
to be relatively shallow, as confirmed by our calculations. 

Then, the traction curves for mode II are derived as numerical derivative of the corresponding sliding energy profiles and
they are presented in the bottom panels of Fig.~\ref{fig7:GSFE}. For the basal sliding, namely 
the slips along $\langle\bar{1}2\bar{1}0\rangle_\mathrm{ZrB_2}$ and $\langle10\bar{1}0\rangle_\mathrm{ZrB_2}$ of interface I$_\mathrm{Zr-C}$, the shear 
strengths are calculated as 1.47~GPa and 0.74~GPa, respectively. In contrast, those along 
$\langle\bar{1}2\bar{1}0\rangle_\mathrm{ZrB_2}$, $\langle0001\rangle_\mathrm{ZrB_2}$ and 
$\langle11\bar{2}3\rangle_\mathrm{ZrB_2}$ for interface III$_\mathrm{Zr-C}$ are, respectively, 1.15~GPa, 1.13~GPa and 
0.66~GPa. Therefore, the slip system \{10$\bar{1}$0\}$\langle11\bar{2}3\rangle_\mathrm{ZrB_2}$ is prone to be activated 
first. The easy activation of the \{10$\bar{1}$0\}$\langle11\bar{2}3\rangle_\mathrm{ZrB_2}$ slip system was previously 
reported for bulk ZrB$_2$ at over 700$\degree$C \cite{Csanadi2017}. The interfacial sliding along 
$\langle\bar{1}2\bar{1}0\rangle_\mathrm{ZrB_2}$ of I$_\mathrm{Zr-C}$ shows comparable shear strengths 
with those of $\{0001\}_\mathrm{ZrB_2}$ and $\{11\bar{2}0\}_\mathrm{ZrB_2}$ of III$_\mathrm{Zr-C}$, although the latter two 
exhibit slightly lower values.     

In order to trace the the sliding energy profile along the minimum energy path, we perform full relaxation calculations 
using the NEB method. This also allows us to extract some key shear parameters. The energy maximum, namely the 
unstable stacking fault energy, $\gamma_\mathrm{us}$, governs the dislocation nucleation at sites of stress concentrations 
such as at the crack tips. The metastable points (local energy minima) correspond to stable stacking faults with their 
energies, $\gamma_\mathrm{sf}$, determining the dislocation core dissociation, the Peierls stress, the dislocation 
energy, and the primary slip planes. These shear parameters ($\gamma_\mathrm{us}$ and $\gamma_\mathrm{sf}$) 
are summarized in Table~\ref{tab:table1}.

\begin{figure}
\centering
\includegraphics[width=6.5in]{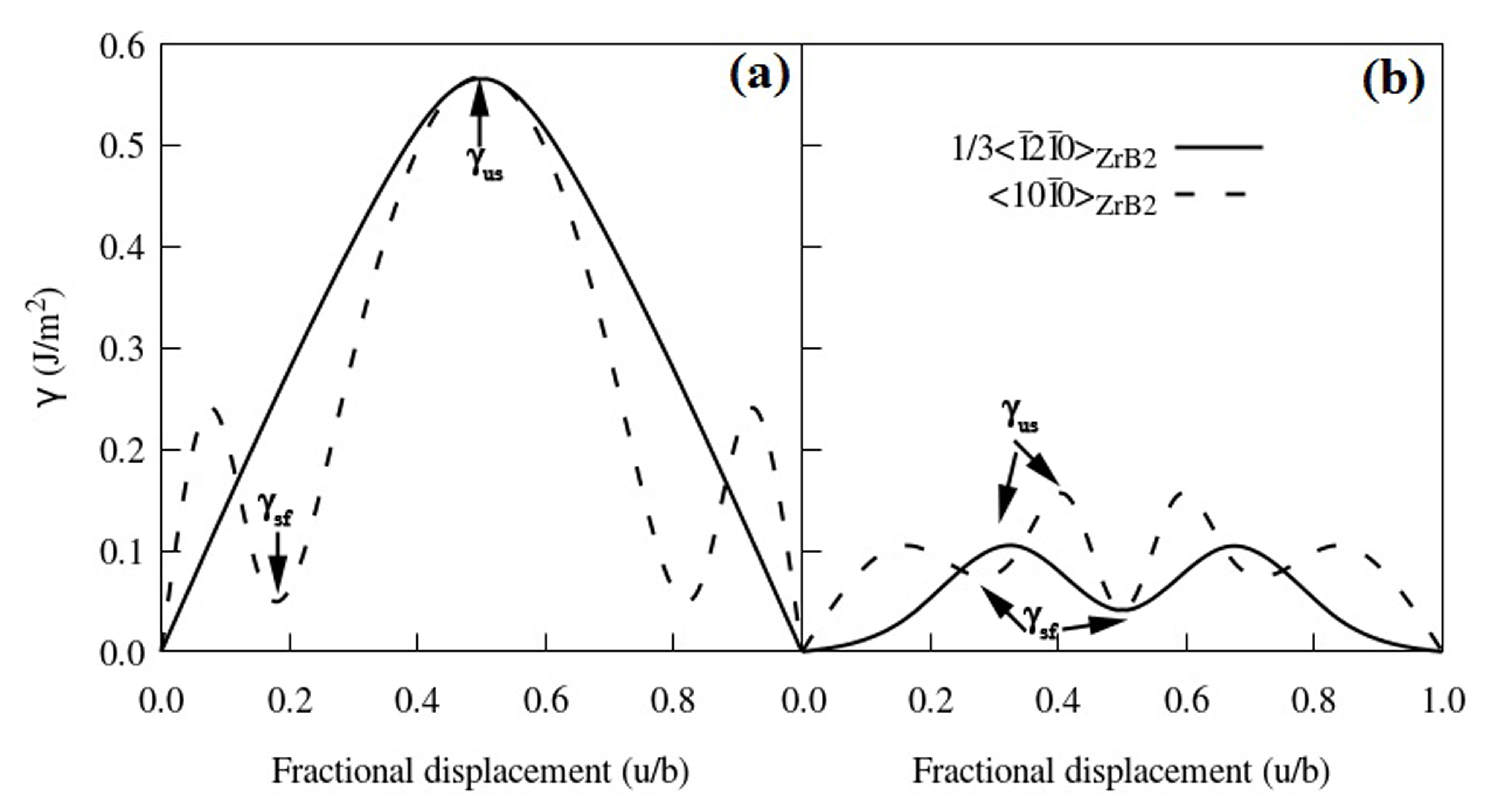}
\caption{Energy profiles of the interfacial sliding obtained by allowing full atomic relaxation (NEB method): 
(a) I$_\mathrm{Zr-C}$ and (b) I$_\mathrm{B-C}$. The  Burgers vectors ($\vec{b}$) are 
$\vec{b}_1$ = $\frac{1}{3}\langle\bar{1}$2$\bar{1}$0$\rangle_\mathrm{ZrB_2}$ and 
$\vec{b}_2$ = $\langle$10$\bar{1}0\rangle_\mathrm{ZrB_2}$.}
\label{fig8:GSFE}
\end{figure}

The relaxed sliding energy profiles are then shown in Figs.~\ref{fig8:GSFE}(a) and \ref{fig8:GSFE}(b) for the interfaces 
I$_\mathrm{Zr-C}$ and I$_\mathrm{B-C}$, respectively. Let us look at the I$_\mathrm{Zr-C}$ case [panel (a)] first. 
We find that $\gamma_\mathrm{us}$ for both the basal $\langle\bar{1}2\bar{1}0\rangle_\mathrm{ZrB_2}$ and 
$\langle10\bar{1}0\rangle_\mathrm{ZrB_2}$ shear is 0.57~J/m$^2$, a value that is almost four times larger than 
those obtained for {\it rigid} shearing. This increase is due to the atomic rearrangement, especially the out-of-plane 
corrugation of the graphene layer, which effectively obstructs the interfacial sliding. In order to understand such 
dramatic increase we note that the total energy reduction due to atomic relaxation differs 
depending on the precise stacking order. In the case of Zr-C interface, the AB stacking corresponds to the initial position of
its sliding energy landscape (see Fig.~\ref{fig7:GSFE}), which shows a deeper energy well after atomic relaxation since some of the 
stress is released. In contrast, the AA configuration is at a peak of the $\gamma$ curve and it is relatively stress-free. 
As a consequence atomic relaxation has little effect on its energetics. Thus, the final result of the relaxation is that of 
increasing the energy barrier, $\gamma_\mathrm{us} = E_\mathrm{AA} - E_\mathrm{AB}$.
This suggests that the r\^ole of corrugated graphene is twofold. On the one hand, it reduces 
the interface adhesion strength during interface debonding, on the other hand it increases the interfacial friction 
during the sliding process.

As for the I$_\mathrm{B-C}$ interface, we note that the $\gamma_\mathrm{us}$ values (0.10-0.15~J/m$^2$) are similar 
to those calculated without performing atomic relaxation. Now $\gamma_\mathrm{us}$ is four times smaller than that of 
I$_\mathrm{Zr-C}$. This is the result of the shallow $\gamma$ surface of the I$_\mathrm{B-C}$ interface, consistent 
with the ranking given before for the adhesive and binding energies. Since the B-C interfaces are bonded by weak physical adsorption (see later discussion), also their 
low energy states have the symmetric AA stacking, rather than the non-symmetric AB or AC ones. Graphene corrugation 
has little effect on the interfacial sliding of B-C interfaces, when compared to the case of Zr-C interfaces. We then conclude
that the B-C interfaces are much more favourable for sliding that the more adhesive Zr-C ones. 

Finally, we compare the ideal shear strength of Zr-C and B-C interfaces with the ZrB$_2$ matrix. The calculated $\tau_\mathrm{c}$ 
for I$_\mathrm{Zr-C}$ are 4.55~GPa and 8.37~GPa, when sliding respectively along 
$\vec{b}_1 = \frac{1}{3}\langle\bar{1}2\bar{1}0\rangle_\mathrm{ZrB_2}$ and 
$\vec{b}_2 = \langle10\bar{1}0\rangle_\mathrm{ZrB_2}$. The same quantities are reduced to 1.87~GPa and 3.35~GPa for the 
corresponding I$_\mathrm{B-C}$ interface and they are increased to 49.62~GPa and 24.45~GPa for bulk ZrB$_2$. 
Thus the interfacial shear strength of ZrB$_2$/graphene interfaces is at least one order of magnitude lower than those of
the ZrB$_2$ matrix and the same observation is valid for the cleavage strength. This suggests that the $\gamma$-surfaces 
are much less corrugated at a heterophase interface, which indicates relatively ease of interfacial sliding. When examining specific interfaces, it is clear that I$_\mathrm{B-C}$ is more prone to host deformation pathways, than I$_\mathrm{Zr-C}$.

\section{Discussion}
\subsection{Interfacial bonding mechanism}\label{4.1}

\begin{figure}
\centering
\includegraphics[width=4.2in]{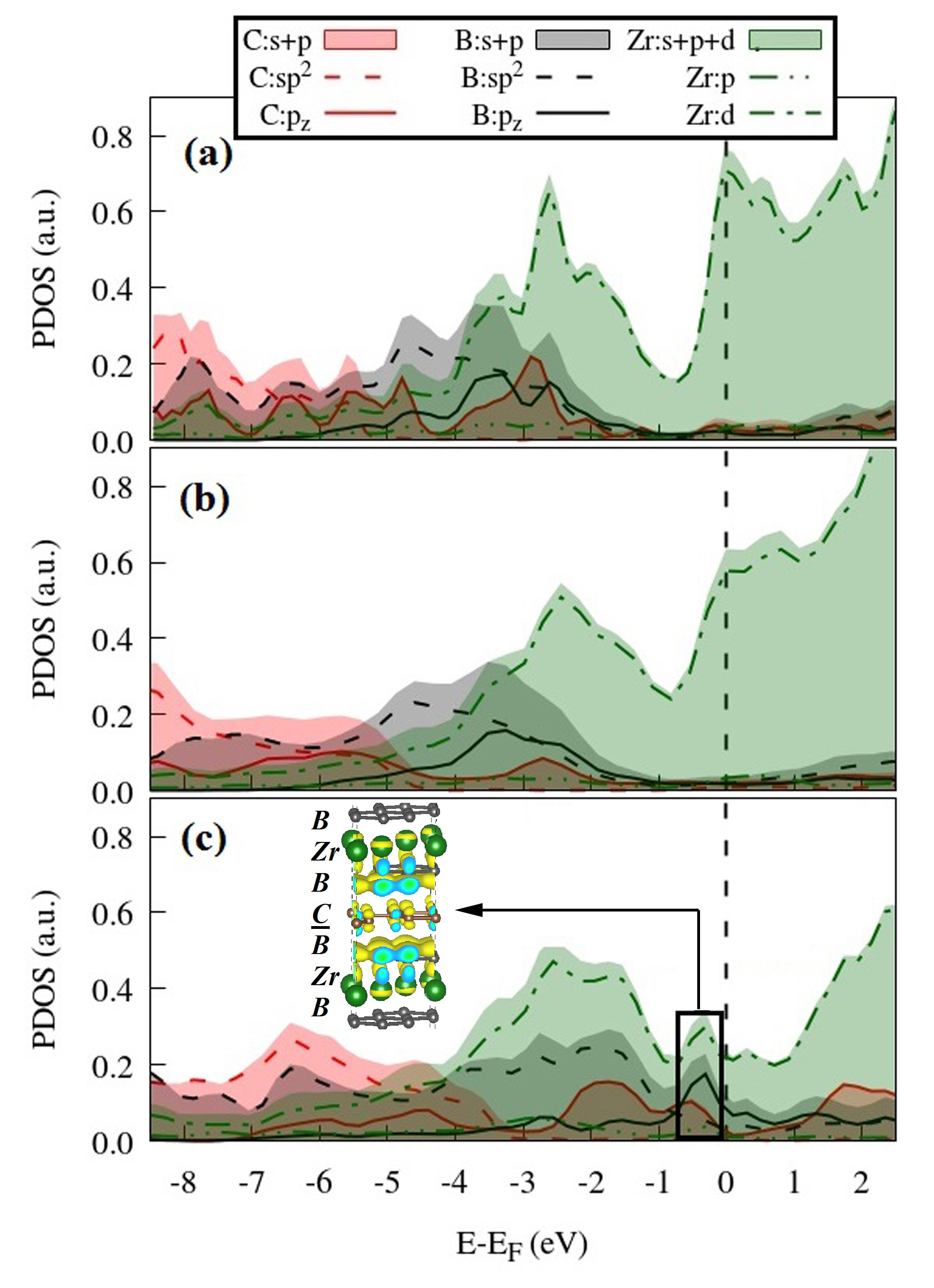}
\caption{(color online) Projected density of states (PDOS) over the C atoms in graphene, and the Zr and B ions in the ZrB$_2$ 
plane next to the graphene sheet for the interface models (a) I$_\mathrm{Zr-C}^\mathrm{AA}$, (b) II$_\mathrm{Zr-C}^\mathrm{AA}$ 
and (c) I$_\mathrm{B-C}^\mathrm{AA}$. The inset in panel (c) shows the projected charge density originating from the 
states corresponding to the $p_z$-orbital peak in the PDOS (the black box).}
\label{fig10:DOS}
\end{figure}

One of the main result from the previous sections is that the Zr-C and B-C interfaces present very different strengths of adhesion 
and binding, as well as the traction curves under the two loading modes investigated. This suggests that different bonding mechanisms are at 
play in these two interfaces. Here, we analyse the interfacial interaction by looking at the projected density of states (PDOSs) and 
the projected charge density, as shown in Figs.~\ref{fig10:DOS}(a) through \ref{fig10:DOS}(c).  

As expected, the partially filled Zr $d$ bands dominate the PDOS for energies starting at 4~eV below the Fermi 
level, $E_\mathrm{F}$, and they represent the main contribution to the Fermi surface for Zr-C interfaces [see 
Fig.~\ref{fig10:DOS}(a) and \ref{fig10:DOS}(b)]. In this case a distinctive feature is the appearance of a 
PDOS reduction at about 1~eV below $E_\mathrm{F}$, which separates a region ($E>E_\mathrm{F}-1$~eV) with little  
contribution from either B or C, from another region ($E<E_\mathrm{F}-1$~eV), where the PDOS from the C 2$p_z$ 
orbital is significant. This is a clear signature of the strong hybridization between the C 2$p_z$ and the Zr 4$d$ 
orbitals, and the consequent formation of Zr-C covalent bonds. Such attribution is consistent with the strong affinity 
between Zr and C and the existence of covalent ZrC$_{1-x}$ compounds. Covalent bonding across interfaces has 
also been suggested for the interfaces of Ti$_2$C/graphene~\cite{Paul2017}, ZrC/SiC~\cite{XIONG2017162} and 
metal graphene contacts~\cite{Giovannetti2008}.

Across the Zr-C interfaces the electron-acceptor B ions (with valence configuration $s^2p^1$) are replaced by 
C ($s^2p^2$). This effectively shifts the Fermi level upwards in energy, and in fact the PDOS pseudogap moves 
from $E_\mathrm{F}$ in bulk ZrB$_2$ to 0.8~eV below $E_\mathrm{F}$ at the Zr-C interface. The effective
charges, as calculated from Bader analysis, of Zr and B in bulk ZrB$_2$ are +1.54$e$ and -0.77$e$
($e$ is the electron charge)~\cite{Wang2013}. These become +1.23$e$ (Zr) and -0.35$e$ (C) at the Zr-C interface,
suggesting that the Zr-C bond has a mixture of ionic and covalent nature. 

The situation for the B-C interfaces is rather different, as one can easily conclude by looking at Fig.~\ref{fig10:DOS}(c). 
In this case the most prominent feature of the DOS is a peak just below the Fermi level, with significant projections originating 
from the B and C 2$p_z$ orbitals. This suggests that the bonding has a $\pi$-$\pi$ stacking nature, a fact that is confirmed
by the projected charge density shown in the inset of Fig. \ref{fig10:DOS}(c). Such $\pi$-$\pi$ interaction is commonly present 
in aromatic compounds~\cite{Neel2017a}. 

\begin{figure}
\centering
\includegraphics[width=4.2in]{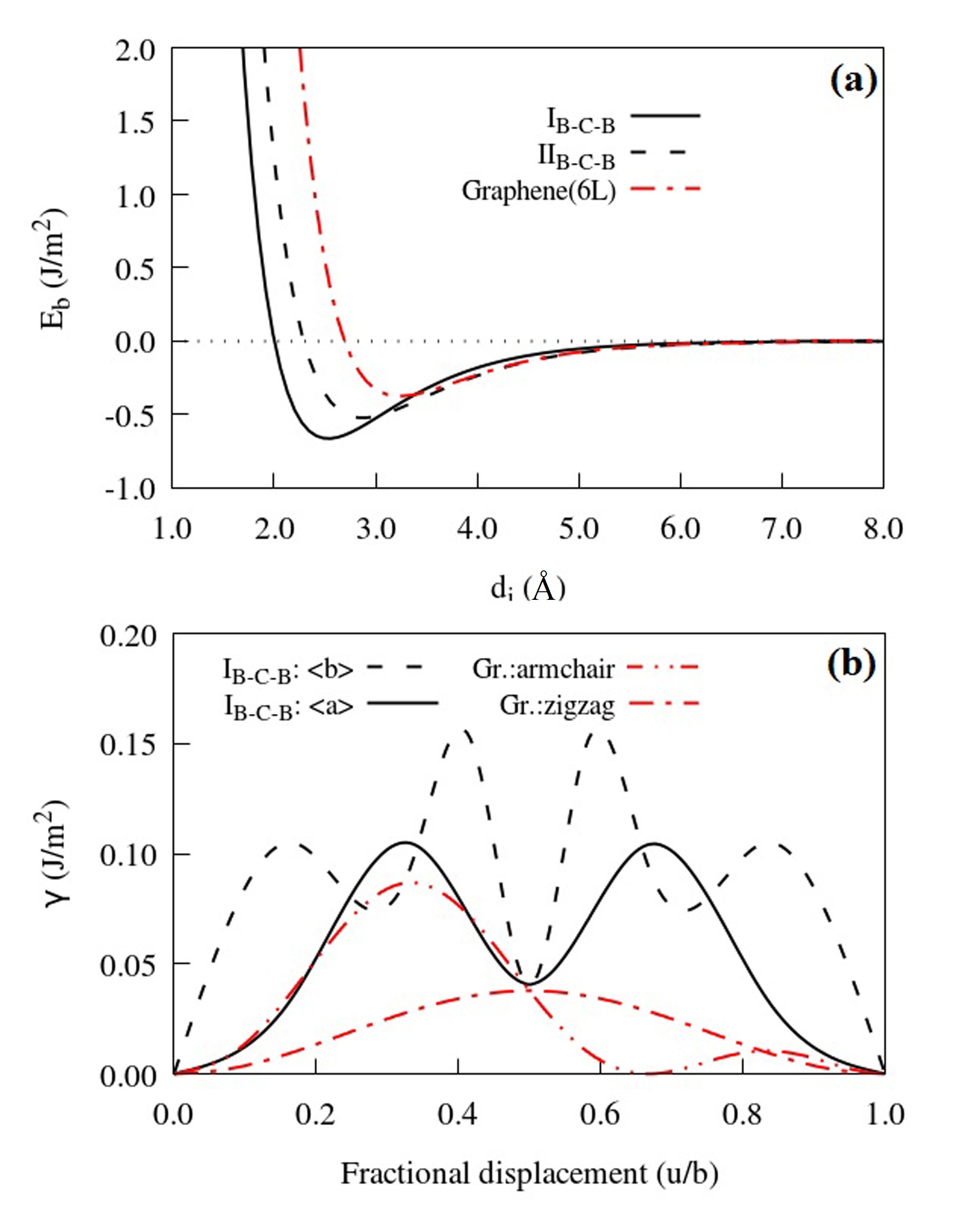}
\caption{Comparison of the mechanical properties of the B-C-B structure with that of graphene. (a) Binding energy, 
$E_\mathrm{b}$ (in J/m$^2$), and (b) sliding energy corrugation, $\gamma$ (in J/m$^2$). In the case of graphene the sliding
modes are along the armchair and zig-zag directions, while for ZrB$_2$/graphene we consider displacements along the 
basal $\langle\bar{1}2\bar{1}0\rangle_\mathrm{ZrB_2}$ and  $\langle10\bar{1}0\rangle_\mathrm{ZrB_2}$ directions.}
\label{fig10:bonds}
\end{figure}

As expected from elementary chemistry the B-C $\pi$-$\pi$ interaction is weaker than the covalent bond between Zr and 
C, as shown by our previous calculations of the mechanical properties of the various interfaces. However, such $\pi$-$\pi$ 
interaction is still stronger than the van der Waals coupling between graphene layers in graphite and, in general, among 
monolayers of van der Waals layered compounds. This is demonstrated in Fig.~\ref{fig10:bonds}(a), where we compare the 
binding energy curves, $E_\mathrm{b}(d_i)$, of the B-C interfaces with that of graphite. The binding energies are calculated 
to be 0.488~J/m$^2$ and 0.663~J/m$^2$ for I$_\mathrm{B-C}$ and II$_\mathrm{B-C}$, respectively. In contrast we compute 
$E_\mathrm{b}(d_0)=0.374$~J/m$^2$ for a 6-layer graphene nanosheet. In addition the equilibrium interlayer distance at the 
B-C interfaces is between 2.5~\AA\ and 3.0~\AA, while that in graphite is 3.23~\AA. 
The $\pi$-$\pi$ stacking interaction here is made stronger than that in graphite by two features of the electronic structure: 
1) the electrostatic interaction between the negatively 
charged C and the positively charged B atoms ; 
2) the offset stacking of the borophene and graphene layers contributing to reduce the repulsive interaction.     
Since the van der Waals interaction among the graphene sheets is believed to be the reason behind its agglomeration in 
composites~\cite{Verma2018}, the stronger adhesion between boraphene and graphene can be exploited as a tool for breaking 
the agglomeration and promoting a more uniform dispersion of graphene nanofillers in ceramic matrices. 
This can be beneficial to the improvements of the overall structural stability and functionality of the nanocomposites. 

Interestingly, the $\pi$-$\pi$ bonding mechanism across the B-C interfaces gives to these interfaces mechanical properties 
analogous to those of 2D materials. This is illustrated in Fig.~\ref{fig10:bonds}(b), where we compare the $\gamma$ sliding 
curves of the two I$_\mathrm{B-C}$ interfaces with those calculated for graphene. We find that the $\gamma_\mathrm{us}$ 
values for graphene are 0.04~J/m$^2$ and 0.09~J/m$^2$, respectively for shearing along the zigzag and armchair edges. 
The slipping along the basal $\langle\bar{1}2\bar{1}0\rangle_\mathrm{ZrB_2}$ and  $\langle10\bar{1}0\rangle_\mathrm{ZrB_2}$ 
directions returns the $\gamma_\mathrm{us}$ values of 0.10~J/m$^2$ and 0.15~J/m$^2$, respectively, which are comparable 
to those of graphene.

\subsection{Interfacial mechanics}

In the previous sections we have observed a large difference  between the Zr-C and the B-C interface with respect 
to their adhesive and mechanical properties. In particular the Zr-C interfaces display stronger adhesion, cleavage and shear 
strength than those of B-C interfaces. This can be well explained by the difference in interfacial bonding mechanism, which is
covalency-dominated in Zr-C and $\pi$-$\pi$-stacking type in B-C interfaces. The stronger Zr-C interfaces thus present high 
resistance to both cracks and interfacial sliding, while the weak B-C interfaces are much more easy to allow interfacial debonding and shear. %

The ZrB$_2$ matrices, have a rich variety of stable surfaces, either Zr-, B- or mix-terminated surfaces with different orientations.
They can be somehow engineered by acting on the ZrB$_2$ growth conditions.  Note the existence of the Zr/B-terminated (0001)
surface has been confirmed by both experiments \cite{0953-8984-20-26-265006,TENGDELIUS201671,doi:10.1116/1.4916565} and theory \cite{Suehara2010,Zhang2018}.
The (0001) surfaces with Zr/B-terminations emphasized here
are the two extreme cases, which exhibit completely different mechanical behavior.
In general, Zr-C interfaces can be prepared by stabilising the Zr-terminated surfaces in Zr-rich conditions. While a B-rich
environment will favour the formation of B-C interfaces. 
Therefore, surface chemistry can be used as an efficient tool for tailoring the interfacial mechanics by tuning the surface contacts. 

It is important to note that the toughening 
effect of graphene fillers is correlated with the interface debonding mechanism. Since in composites the interfacial debonding will 
favour the dissipation of energy via crack bridging, crack deflection and fillers pulling-out, the predominance of a mechanism over 
the others will affect the crack propagation behavior along  various interfaces, so that different ZrB$_2$ growth conditions may result 
in tuning possibility of the mechanical properties of GCMCs. This observation also provides hints on why there is a wide variation of 
the fracture-toughness-parameter values in experiments on graphene-reinforced composites~\cite{Miranzo2017}. 

Last but not least, monolithic ZrB$_2$ and graphene nanoplatelets are used as starting materials during spark plasma
sintering~\cite{Yadhukulakrishnan2013} of ZrB$_2$/graphene composites. The nanoplatelets are made of short 
stacks of ribbon-shaped graphene sheets. These are functionalised with groups like ethers, carboxyls or hydroxyls, which may further 
modify the interfacial bonding~\cite{Zhang2018-He}. Therefore, a scrupulously designed surface treatment is vital for the accurate
control of the interfacial interaction, and hence of the interface mechanics and the failure mode. Our theoretical results suggest that
the preferred interface mechanics can be engineered by acting on the interfacial interactions through the tuning of the surface chemistry 
of ZrB$_2$ and graphene.

\section{Conclusion} 

In this work we have studied the heterophase interfaces of graphene-reinforced ZrB$_2$ composites based on DFT 
calculations. A number of atomistic models for the various possible interfaces has been constructed by using the most 
thermodynamically stable surfaces of the ZrB$_2$ matrix. Then, we have systematically investigated their interface 
adhesion, mechanical behaviour and bonding mechanism. We have demonstrated that two kinds of interfaces, namely 
Zr-C-Zr and B-C-B, offer a wide spectrum of mechanical properties due to their dissimilar interaction between the constituent 
materials. In particular Zr-terminated surfaces bind to graphene in a covalent way, while the interaction with the borophene 
planes in ZrB$_2$ has a $\pi$-$\pi$-stacking nature.
By tuning the surface chemistry of the ZrB$_2$ matrix one can prepare composites that expose graphene predominantly to 
a specific ZrB$_2$ termination (either Zr or B or mixed), so it is possible to go from a regime of weak graphene/matrix interaction to
one where the interaction is relatively strong. This provides an important design scheme in the synthesis of ceramic composites. 
Furthermore, the fact that the borophene/graphene interface is more adhesive than the graphene/graphene one may offer 
an opportunity to tackle  flakes agglomeration, a issue encountered when processing most graphene-reinforced composites.
Finally, we have analysed the r\^ole of graphene rippling over the mechanical properties of the interfaces. Interestingly we 
have found that rippling drastically increases the friction for sliding graphene over ZrB$_2$ in the case of Zr termination,
while it has no significant effect for that with B termination.



\begin{acknowledgement}
This work is supported by the European Union's Horizon 2020 ``Research and innovation programme'' under the grant agreement 
No. 685594 (C$^3$HARME). Computational resources have been provided by the Irish Center for High-End Computing (ICHEC) 
and the Trinity Centre for High Performance Computing (TCHPC). The authors also thank Rui Dong for the {\sc clattice} code used
to merge surface slabs. 

\end{acknowledgement}

\begin{suppinfo}

The Supporting Information is available free of charge on the ACS Publications website for:
Geometry details of interface models, Structural relaxation effects on interfacial cleavage.

\end{suppinfo}

\providecommand{\latin}[1]{#1}
\makeatletter
\providecommand{\doi}
  {\begingroup\let\do\@makeother\dospecials
  \catcode`\{=1 \catcode`\}=2 \doi@aux}
\providecommand{\doi@aux}[1]{\endgroup\texttt{#1}}
\makeatother
\providecommand*\mcitethebibliography{\thebibliography}
\csname @ifundefined\endcsname{endmcitethebibliography}
  {\let\endmcitethebibliography\endthebibliography}{}


\end{document}